\documentclass[fleqn,10pt]{manuscript}
\usepackage{soul}
\usepackage{graphicx}%
\usepackage{url}
\usepackage{multirow}%
\usepackage{amsmath,amssymb,amsfonts}%
\usepackage{amsthm}%
\usepackage{mathrsfs}%
\usepackage[title]{appendix}%
\usepackage{xcolor}%
\usepackage{textcomp}%
\usepackage{manyfoot}%
\usepackage{booktabs}%
\usepackage{algorithm}%
\usepackage{algorithmicx}%
\usepackage{algpseudocode}%
\usepackage{natbib}
\usepackage{rotating}
\usepackage{tabularx}
\usepackage{threeparttable}
\usepackage{pdflscape}
\usepackage[T1]{fontenc}
\usepackage[utf8]{inputenc}

\title{Big Bang Nucleosynthesis results refined via the Trojan Horse Method}

\author[1,2]{Roberta Spartà}
\author[2,3]{Rosario Gianluca Pizzone}
\author[2,3,4]{Livio Lamia}
\author[2]{Alessandro Alberto Oliva}
\author[2]{Marco La Cognata}
\author[2]{Alessia Di Pietro}
\author[2]{Pierpaolo Figuera}
\author[2]{Giovanni Luca Guardo}
\author[5,6]{Marco La Commara}
\author[1,2]{Dario Lattuada}
\author[7,8]{Marco Mazzocco}
\author[9,10]{Sara Palmerini}
\author[2,3]{Giuseppe Gabriele Rapisarda}
\author[2,3,4]{Stefano Romano}
\author[2,3]{Maria Letizia Sergi}
\author[1,2]{Aurora Tumino}

\affil[1]{Dipartimento di Ingegneria e Architettura, Università di Enna 'Kore', Enna, Italy}

\affil[2]{Laboratori Nazionali del Sud, INFN, Catania, Italy}

\affil[3]{Dipartimento di Fisica e Astronomia ‘Ettore Majorana’, Universitá di Catania, Catania, Italy}

\affil[4]{Centro Siciliano di Fisica Nucleare e Struttura della Materia (CSFNSM), Catania, Italy}

\affil[5]{Dipartimento di Farmacia, Università degli Studi di Napoli "Federico II", Napoli, Italy}

\affil[6]{Sezione di Napoli, INFN, Napoli, Italy}

\affil[7]{Dipartimento di Fisica e Astronomia 'Galileo Galilei',Universitá degli Studi di Padova, Padova, Italy}

\affil[8]{Sezione di Padova, INFN, Padova, Italy}

\affil[9]{Dipartimento di Fisica e Geologia, Universitá degli Studi di Perugia, Perugia, Italy}

\affil[10]{Sezione di Perugia, INFN, Perugia, Italy}

\begin{abstract}
\hl{This work presents the Trojan Horse Method (THM) as a powerful technique for measuring nuclear reaction cross sections at astrophysical energies. We then explore the impact of THM-derived reaction rates on the predictions of Standard Big Bang Nucleosynthesis (SBBN) using the PRIMAT code. Primordial abundances are shown for the single rate impact and, for the first time, also for all the THM rates together. 
The result shows significant differences with the use of THM rates, which in some cases goes in the direction of improving the agreement with the observations with respect to the use of only reaction rates from direct data, especially for the $^7$Li and deuterium abundances, which are still open issues for SBBN.}
\end{abstract}
\begin{document}

\flushbottom
\maketitle
\thispagestyle{empty}

\section*{Introduction}\label{sec1}

The use of indirect methods in experimental nuclear astrophysics allows the success in most of the low energy reactions challenges. The first among these is the presence of the Coulomb barrier, as the positive charges of nuclei repel one another, that exponetially reduces the fusion cross sections in astrophysical plasma. In this sense, indirect methods have prevented the use of extrapolation of the cross section $\sigma(E)$ at low energies from higher energies (typically $E>100$ keV), that can introduce additional uncertainties due to potentially unobserved resonances at low energies.

Indirect approaches have been developed via direct reaction mechanisms such as transfer (stripping and pickup) and quasi-free (QF) knockout reactions. One particularly successful method is the Trojan Horse Method (THM), designed to extract the bare nucleus cross section $\sigma_b(E)$ of binary interactions at astrophysical energies by studying three-body reactions- The $\sigma_b$ is {\em bare} because the THM bypasses both Coulomb barrier suppression and electron screening shielding effects, yielding direct access to bare nucleus cross sections without reliance on potentially unreliable extrapolations.
 
A special astrophysical environment where THM has proved its attainment is Big Bang Nucleosynthesis (BBN). As one of the earliest processes in the universe, BBN explores epochs of few minutes after the Big Bang, encompassing temperatures below 1 MeV. This timeframe is fundamental for bridging Cosmology to Nuclear Physics. The Standard BBN (SBBN) model predicts the formation of light nuclei -such as $^2$H, $^3$He, $^4$He, and $^7$Li - derived from initial protons and neutrons. 
Observations today show reasonable agreement with SBBN predictions in the relevant astrophysical contexts for these isotopes, but for $^3$He and $^{7}$Li. The $^3$He abundance is difficult to measure and only upper limits are given \cite{BalserBania}. The $^7$Li abundance is predicted to be three times the observed value in stars and such a mismatch is usually know as the Cosmological Lithium Problem, $CLiP$. This general accordance, particularly when comparing primordial abundances derived from Cosmic Microwave Background (CMB) anisotropy from the Planck probe data with theoretical predictions, helps in constraining the baryonic density $\Omega_b$, a critical parameter in the SBBN framework.
The latest result from Planck \cite{aghanim} is also the value adopted in our current calculations and is $\Omega_{b}h^2$=0.02242$\pm$0.00014 (where $h$ is the Hubble parameter).
The accuracy of SBBN predictions calls for reliable nuclear reaction rates, which require precise measurements of cross sections at relevant energies. 
Twelve primary processes dominate the SBBN network, as addressed in sec. \ref{sec2}. 
Moreover, the precise knowledge of low energy cross section of many of the reactions involved in such a network is essential to address the $^7$Li observations challenge, because it is also depleted in stars \cite{Sbordone2010}. Ideally, these cross sections should be measured within the astrophysical Gamow window, typically at a few tenths or hundreds of keV \cite{Iliadis}. Efforts have been made to perform direct measurements at these energies, often in underground laboratories (e.g. \cite{LUNA}) or using ultrapure targets and/or intense beams. 

Nonetheless, many open questions have still to be solved, the first being the above mentioned $^3$He and $^7$Li disagreement of predictions with observations. CLiP has also been associated to stellar evolution \cite{Fu},\cite{Korn2006} and cosmological \cite{Mathews2020} causes, beyond the nuclear ones (reaction rates uncertainties). Recently, a measurements of d(p,$\gamma$)$^3$He \citep{Mossa2020} has opened questions on the deuterium primordial abundance, claiming D/H=(2.439$\pm$0.037)$\times$10$^5$, noting a 1.8$\sigma$ difference with respect to the value suggested by the Cosmic Microwave Background (CMB) and the baryonic oscillation data \cite{anewtension}.

To answer the open questions of BBN, we measured the cross sections of six out of the twelve reactions of the main network of BBN (see sec. \ref{sec2}). Primordial abundances calculated with the four BBN reaction rates from THM measurements can be found in \cite{Sparta2021}. To evaluate the single effect of each of the six THM measurements and to implement the $^{3}$He(n,p)$^3$H \cite{Pizzone3He} and $^{7}$Be(n,p)$^{7}$Li \cite{Hayakawa2021} in the overall THM impact on SBBN, we performed the present calculations using the PRIMAT code \cite{Pitrou2020}.

\section{The Trojan Horse Method} \label{sec2}

\begin{figure}
    \centering
    \includegraphics[width=0.5\linewidth]{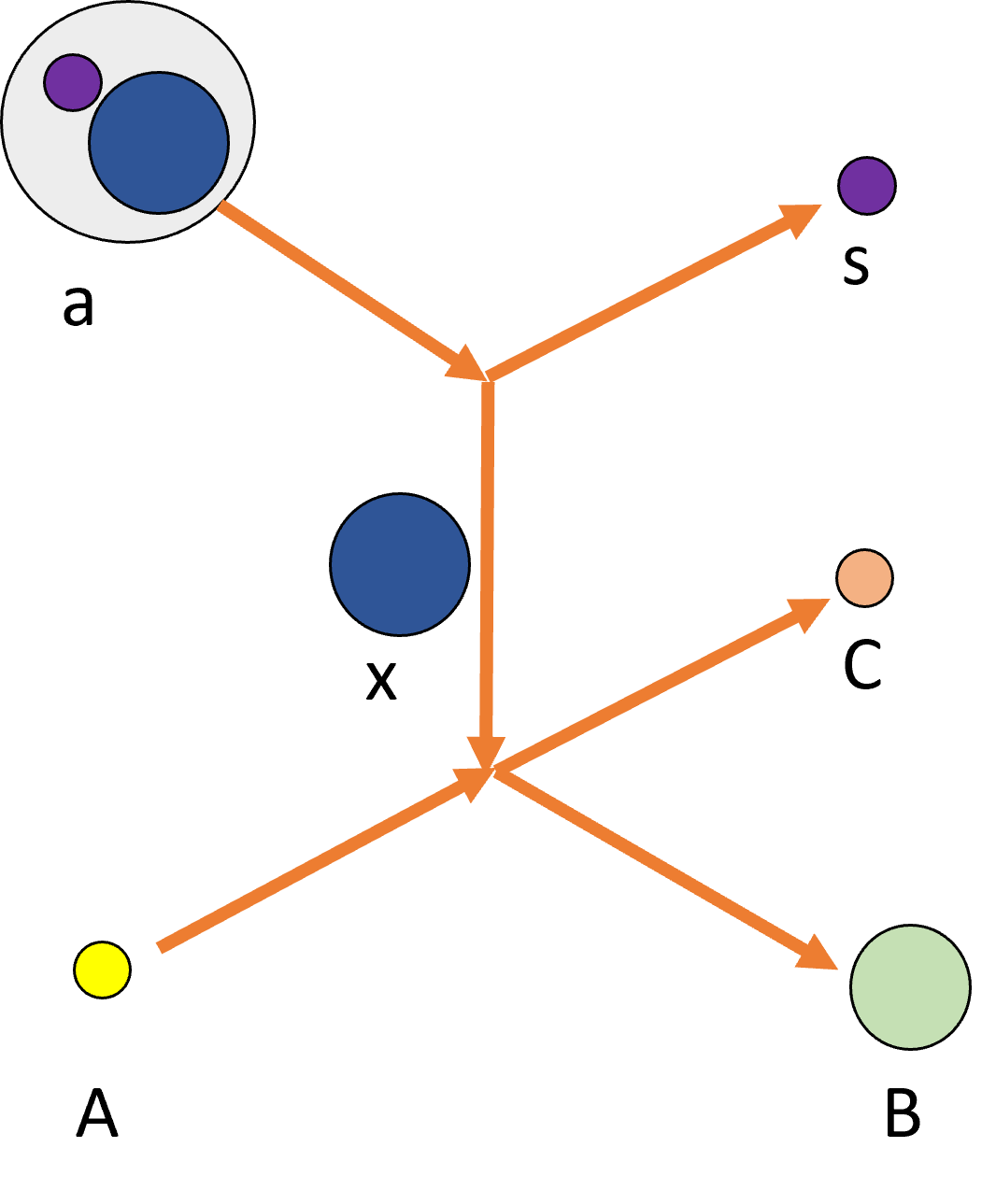}
    \caption{Sketch of the mechanism to be investigated with THM. The $a$ nucleus breaks up in the two clusters $s$ and $x$, the spectator and the participant to the binary reaction of astrophysical interest with $A$, respectively.}
    \label{THM}
\end{figure}

In the THM the reaction of interest $A + x \rightarrow B + C$ is embedded in a three-body breakup reaction of the form $a + A \rightarrow s + B + C$ (sketch in fig. \ref{THM}), where $a = (s + x)$ is the Trojan horse particle, $s$ is the spectator particle, and $x$ represents the actual projectile. Because THM rests upon the selection of the quasi-free reaction mechanism, in this regime the spectator particle $s$ remains unaffected by the interaction, while the particle $x$ participates in the binary reaction of interest with nucleus $A$.

The selection of the quasi-free reaction mechanism enables the extraction of the reaction cross section for the binary system at astrophysical energies by analyzing the energy and angular distributions of the reaction products. 

Further details and formulas can be found in \citep{Spitaleri2019} and \cite{Tumino2021}, for example on how the Trojan Horse nucleus binding energy is compensating the projectile energies, allowing to reach low energies in the c.m. and how it is possible to span a wide region of the excitation function with only one beam energy, selecting a portion of the spectator momentum where the quasi-free mechanism is favored. This allows to cover the whole astrophysically interesting region of the excitation function with a single measurement, thus limiting discrepancies between the many data sets of literature, with their systematic errors.

THM was first used for investigating reactions among charged particles, when nuclear fusion processes are hindered by the Coulomb barrier, and later successfully applied to neutron-induced reactions \cite{Sparta10B},\cite{neutrondriven},\cite{Gulino2010},\cite{Pizzone3He},\cite{Hayakawa2021}. The method has also been instrumental in investigating several reactions of astrophysical significance complicated by the involving Radioactive Ion Beams (RIBs)   \cite{Belicos},\cite{Pizzone18F}. 

Recently, the THM applications were extended to include reactions involving heavier nuclei as for the $^{12}$C+$^{12}$C fusion \cite{TuminoNature}, which are relevant for nucleosynthesis in massive stars. The method has been tested rigorously to confirm its applicability, examining invariance under various experimental conditions, including target/projectile breakup invariance \cite{Musumarra2001} and spectator invariance \cite{Tumino2006}\cite{Pizzone2011}\cite{Pizzone2013}. 
For these reasons, THM has become an essential tool for deriving astrophysical reaction rates, presenting a robust and reliable complement to direct measurements. 

\section{Using THM to shed light on BBN}

\begin{figure}
    \centering
    \includegraphics[width=0.7\linewidth]{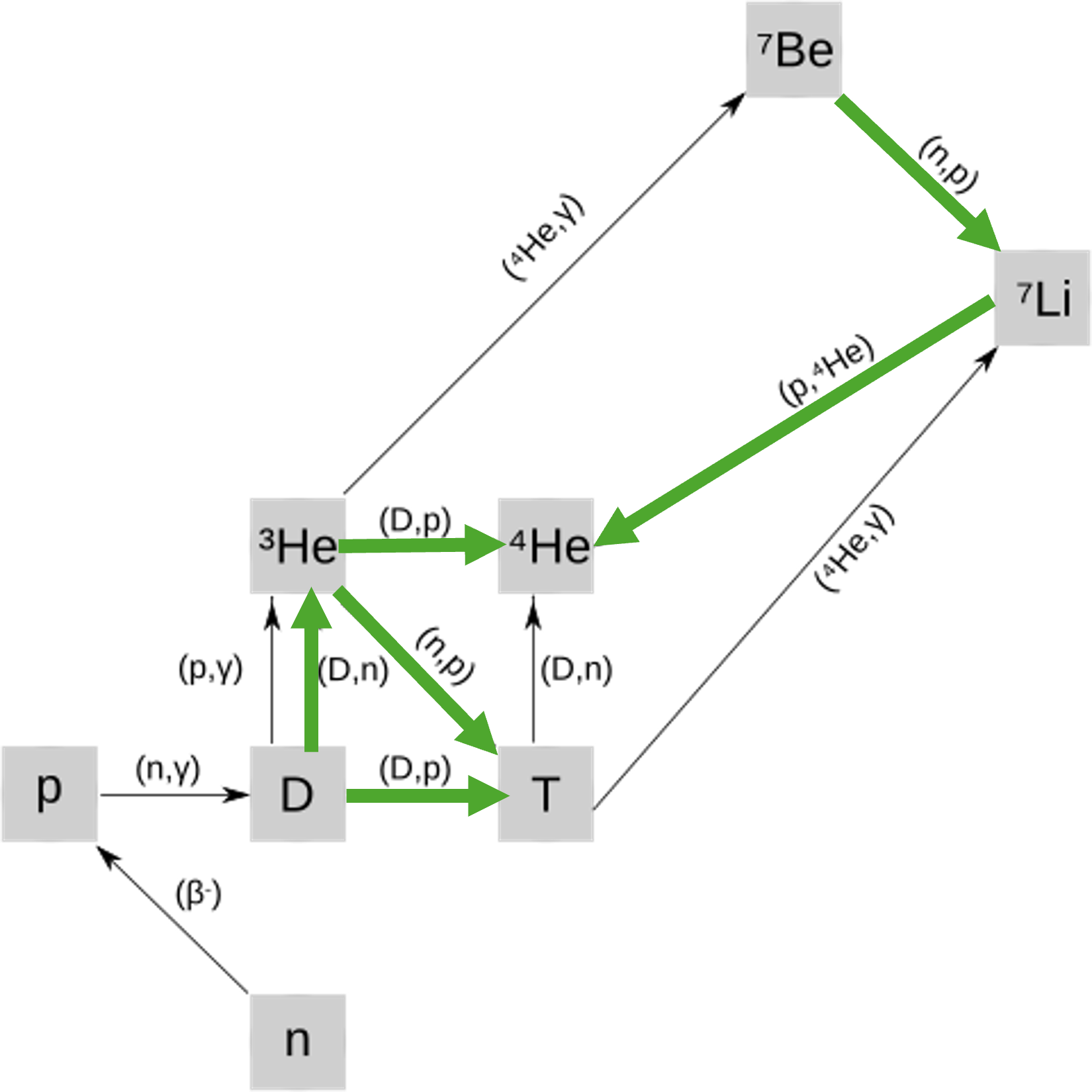}
    \caption{The main reactions network for the SBBN model. The six out of the twelve measured with THM are indicated by green arrows.}
    \label{network}
\end{figure}

In fig. \ref{network} the network of the most influent twelve reactions for the BBN outcomes is shown. The six ones measured with THM are marked with a green arrow. 
To implement the THM reaction rates into the Big Bang Nucleosynthesis scenario, we used the PRIMAT code \cite{Pitrou2018} (version 0.2.2 of 2023).
PRIMAT is based on the \textit{Mathematica} platform and the new version has many updates, such as the d(p,$\gamma$)$^3$He rate in \cite{Moscoso} for T$_9$ $\le$ 4 and from \cite{Coc2015} for 4 $<$ T$_9$ $<$ 10 (being T$_9$=10$^9$K).

Results will be discussed in sec. \ref{res}.

\subsection{Adopted reaction rates}

In tab. \ref{table:sources} we list the references for the THM data used and the corresponding reaction rates from direct data used in PRIMAT. Reaction rates from \cite{Gomez2017},\cite{deSouza2019} and \cite{deSouza2020} are based on bayesian reanalysis of the literature data sets, while \citep{Desc2004} uses R-matrix fits of the available S-factors to calculate the rates.
The THM was successfully applied to investigate $d+d$ reactions in \cite{Tumino14}, enabling the extraction of cross section data over a remarkably broad energy range, exceeding 1 MeV. This comprehensive dataset fills the gaps between direct measurements and extends down to 2 keV in the excitation function.
The reaction rates used for calculations come from the analytical rate form in \cite{Pizzone2014}.
The mirror nuclei reaction $^3$He(n,p)$^3$H was measured with the deuteron as Trojan Horse nucleus, to avoid experimental issues present in all the n-induced direct reactions. Results were published in \cite{Pizzoneepj3He} and, recently, in \citep{Pizzone3He}.
$^3$He(d,p)$^4$He measurement details can be found in \cite{LaCognata3Hedp} and the TH reaction rate can be found in \cite{Pizzone2014}. 
$^7$Li(p,$\alpha$)$^4$He measurement is described in \cite{Lattuada2001} and the reaction rate used can be found in \cite{LamiaAA}. $^7$Be(n,p)$^7$Li measurement and the reaction rate are present in \cite{Hayakawa2021}, where also this reaction rate application to PRIMAT (first version) is shown. 

\begin{table}
    \centering
    \begin{tabular}{ccc}
         \textbf{Reaction}&  \textbf{TH data}& \textbf{PRIMAT data}\\
        $^2$H(d,n)$^3$He&  \citep{Tumino14}&  \citep{Gomez2017}\\
        $^2$H(d,p)$^3$H&  \citep{Tumino14}& \citep{Gomez2017}\\
        $^3$He(n,p)$^3$H&  \citep{Pizzone3He}& \citep{Desc2004}\\
        $^3$He(d,p)$^4$He&  \citep{LaCognata3Hedp}& \citep{deSouza2019}\\
        $^7$Li(p,$\alpha$)$^4$He& \citep{LamiaAA} & \citep{Desc2004}\\
        $^7$Be(n,p)$^7$Li&  \citep{Hayakawa2021}& \citep{deSouza2020}\\
         &  & \\
    \end{tabular}
    \caption{References for the reaction rates used in the calculations in sec. \ref{res}.}
    \label{table:sources}
\end{table}

\section{Results} \label{res}

We used the THM reaction rates as input in PRIMAT separately, replacing, in turn, one by one the direct ones with the THM ones and then we replaced all the rates together, with the results shown in fig.\ref{plot1}. Also the PRIMAT abundances with only direct reaction rates are shown as filled circles for comparison. Observations are shown, where available, as solid lines, together with their error as colored bands. References used are \cite{Cooke} for deuterium, \cite{Bania2002} for $^3$He measured in the Milky Way (red band) and from \cite{Cooke2022} in the Orion Nebula (blue band), \cite{Aver} for $^4$He, \cite{Sbordone2010} for $^7$Li. 

Results of PRIMAT with only direct reaction rates (filled circles) and the ones with all the six THM reaction rates (reverse triangles) have uncertainties calculated with the method described in \cite{Coc2014}, considering lognormal distributions for the probability density functions of the reaction rates (or equivalently their logarithm as Gaussian). Error bands shown are the 15\% and 86\% quantiles of the primordial abundances distributions, obtained with MonteCarlo calculations.
For the present calculations with the single THM rates, we assume the same percentage errors of the reaction rates changed. 

$^6$Li, $^9$Be, $^{11}$B and CNO results are listed in tab. \ref{tab:calc}. PRIMAT column shows the direct reaction rates result, then all the single reactions change are listed from the second to the eight columns, finally the last two columns are for the overall THM rates result and the values calculated in  \cite{Coc2012}, for comparison, being very difficult to get these isotopes abundances with observations.

\subsection{Deuterium, $^{3,4}$Helium and $^{7}$Li primordial abundances}
\begin{figure}
    \centering
    \includegraphics[width=1\linewidth]{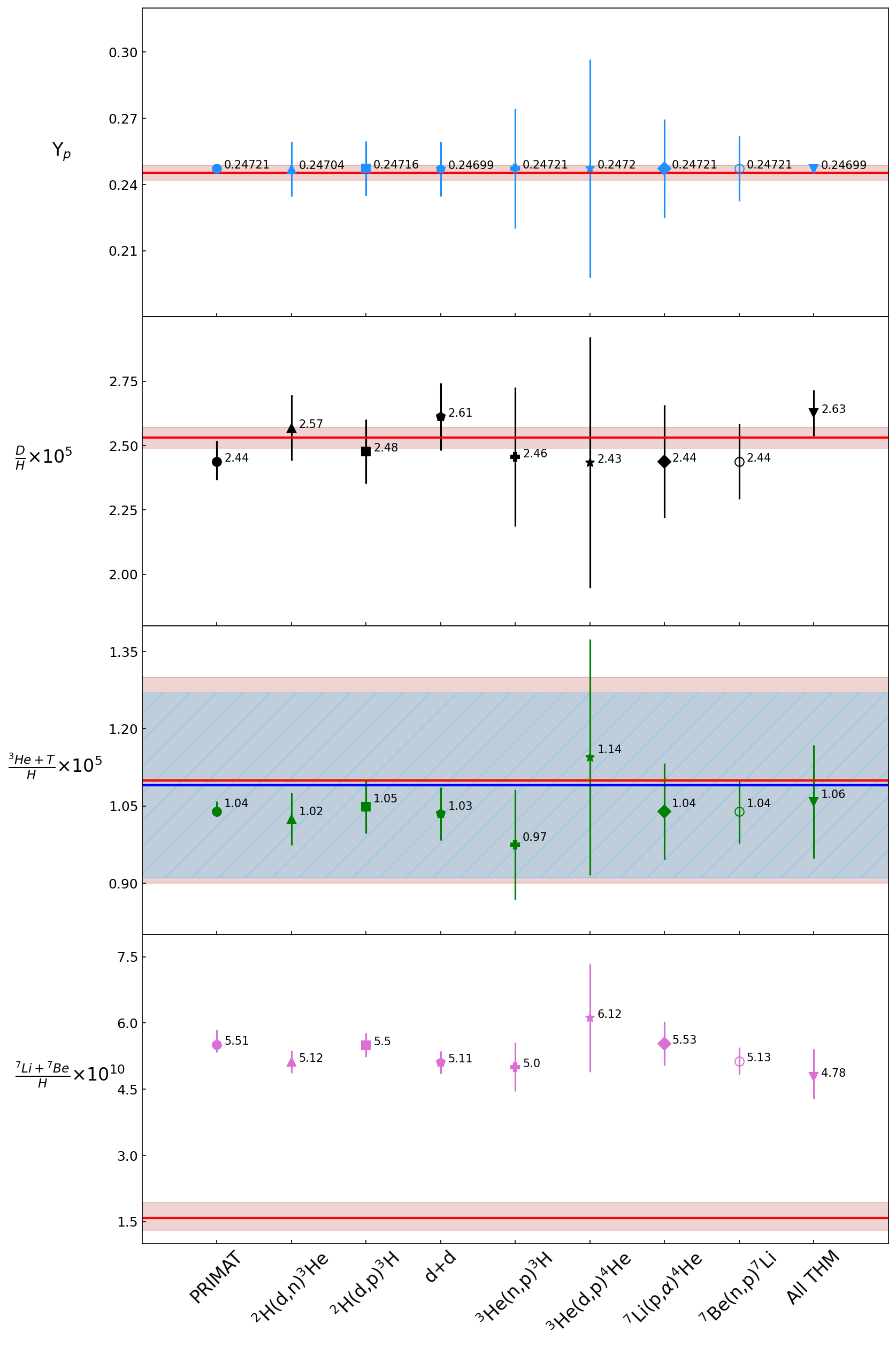}
    \caption{BBN abundances resulting from PRIMAT with only direct data reaction rates (PRIMAT column of points), compared with the PRIMAT results using the THM rates of the six reactions measured (sec. \ref{sec2}) and all together (reverse triangles). Colored bands are the values of reference from observations (Y$_p$ \cite{Aver}, D/H \cite{Cooke},($^3$He+T)/H \cite{BalserBania} in blue and \cite{Cooke2022} in red, ($^7$Li+$^7$Be)/H \cite{Sbordone2010}.}
    \label{plot1}
\end{figure}
The $^4$He abundance, which is usually expressed as the mass fraction Y$_p$ and is tightly linked to the Universe rate expansion, is leveraged by the $d+d$ THM rates (blue triangle, square and pentagon), bringing the primordial prediction of all the THM reaction rates together to Y$_p$=0.24699$^{+0.00005}_{-0.00006}$, blue reverse triangle of fig. \ref{plot1}), which is still inside the observed value by \cite{Aver} of Y$_p$ = 0.2453 $\pm$ 0.0034. The 4$\sigma$ difference between indirect and direct result claims for further investigations. 

The $d+d$ reactions with THM (black pentagon of fig. \ref{plot1}) raises $\frac{D}{H}$$\cdot$10$^5$ to 2.61$\pm$ 0.13, and this reflects into the combined effect of all the THM measurements (black reverse triangle), where the deuterium abundance is 7.7 \% more than the direct result. This increase reverse and confirm the tension between predicted and observed values advanced in \cite{anewtension} (see sec. \ref{sec1}). 

$^3$He and Tritium (eventually decaying into $^3$He) abundance is affected by THM rates changing, but remains in the observational references, which are measured to be $\frac{^3He}{H}*10^{5}$ = 1.1$\pm$0.2 \cite{Bania2002} in the Milky Way and  $\leq$ 1.09$\pm$0.18 \cite{Cooke2022} in the Orion Nebula (see in fig. \ref{plot1} the red and blue bands, respectively). 
In particular, the lower value for the primordial $^3$He abundance obtained with the THM $^3$He(n,p)$^3$H rate ($\frac{^3He+T}{H}$$\cdot$10$^5$=0.97$\pm$0.10, green cross) is compensated by its increase by the THM $^3$He(d,p)$^4$He ($\frac{^3He+T}{H}$$\cdot$10$^5$=1.14$\pm$0.22, green star), resulting in a final abundance (
green reverse triangle) very close to the direct reaction rates result (green filled circle).

The $^7$Be(n,p)$^7$Li (purple empty circle), $^2$H(d,n)$^3$He (purple triangle) and $^3$He(n,p)$^3$H (purple cross) THM measurements have an impressive impact in lowering the $^7$Li final abundance, considering the eventual decay of $^7$Be, thus on the CLiP, reducing the discrepancy with observations, $\frac{^7Li}{H}*10^{10}$ = 1.58$^{+0.35}_{-0.28}$ \cite{Sbordone2010}. 
The combined effect (purple reverse triangle) of these three reaction rates ends up in a 13\% reduction of the A=7 nuclei abundance, with respect to the abundance from only direct data. 
It is worth noticing then how important is to have precise measurements in order to reduce the THM rates uncertainties that proved to be really helpful in decreasing the tension.

A CLiP possible solution through the $^6$Li depletion is thoroughly analyzed in \cite{Fields2022}. This solution and this new value from THM can have the final say on the CLiP, bringing the SBBN model to a further level of concordance.

Light elements primordial abundances were also calculated in  \cite{Pizzone3He} with a previous version of PRIMAT, by using only the $^3$He(n,p)$^3$H rate, leading to  ($\frac{^7Li+^7Be}{H}$$\times$10$^{10}$ =  5.134.
The same holds when considering the calculation in \cite{Hayakawa2021}, where that previous version of PRIMAT was used to reveal the $^7$Be(n,p$_0$)$^7$Li and $^7$Be(n,p$_1$)$^7$Li THM rates on $\frac{^7Li+^7Be}{H}$$\times$10$^{10}$ abundance. The channels sum $^7$Be(n,p$_0$+p$_1$)$^7$Li rate provided an abundance of 5.18$\times$ 10$^{-10}$. 
This above mentioned abundance turns out to be 4.78$\times$ 10$^{-10}$ when using all the THM rates in the calculation (purple reverse triangle), therefore the comprehensive use of the THM data further strengthens the method impact on the CLiP solution.

\subsection{$^6$Li, $^9$Be, $^{11}$B}
For these isotopes, abundance are obtained with PRIMAT without an error estimate, that will be object of forthcoming calculations, and thus are presented in tab. 2. Reliable values of their primordial abundances cannot be easily inferred by observations, thus their reference values for comparison in tab. \ref{tab:calc} are taken from \cite{Coc2012}, calculated at WMAP7 baryonic density, $\Omega_b h^2$=0.02249$\pm$0.00056 
\cite{Komatsu2011}, and using reaction rates obtained with the TALYS code \cite{Goriely2008} (uncertainties are given in \cite{Coc2014}). For all of these nuclei, THM rates bring their values closer to the \cite{Coc2012} predictions.

In particular, $^6$Li is predicted to be more abundant, meeting the prediction in  \cite{Coc2012},  but still very far from the upper limit of  $^6$Li/H =  10$^{-11}$ (observed in the HD84937 star \cite{Steffen}), since the $^6$Li plateau is now known to not exist \cite{Fields2022}. The THM $^2$H(d,n)$^3$He increases the $^6$Li/H value by 5\% (second column in tab. \ref{tab:calc}), but the overall use of the THM rates increases it of 9 \%.

The same holds for the $^9$Be, whose abundance from the model is higher with respect to the one resulting from non-THM rates (PRIMAT column of tab. \ref{tab:calc}), approaching the abundance upper limit observed in metal poor stars ([Fe/H] = - 3.5) as $\frac{^9Be}{H}$ = 3$*$10$^{-14}$.

The corresponding limit for $^{11}$B, namely the lowest boron abundance measured in the galactic halo ([Fe/H] $<$ - 3), gives 10$^{-12}$, while the SBBN codes output is typically 3$\times$10$^{-16}$. The calculated abundance with all the THM rates is now lower than the direct reaction rates (third line of tab. \ref{tab:calc}). 

\subsection{CNO}
The CNO isotopes primordial abundance can affect Population III stars evolution also at abundances $\frac{CNO}{H}$$\sim$10$^{-11}$- 10$^{-13)}$. 
In \cite{Coc2014} it is clearly shown how the tails of the probability distribution overlap the 10$^{-13}$ limit, hence the values predicted with THM reaction rates confirm the scenario in which Population III stars are affected by the primordial production of CNO isotopes, still remaining in a Standard BBN context. 

\begin{sidewaystable}
\centering
\begin{tabular}{|c|c|c|c|c|c|c|c|c|c|c|} \hline 
 & {\bf PRIMAT} & {\bf $^2$H(d,n)$^3$He} & {\bf $^2$H(d,p)$^3$H} & {\bf d+d} &  {\bf $^3$He(n,p)$^3$H} & {\bf $^3$He(d,p)$^4$He} & {\bf $^7$Li(p,$\alpha$)$^{4}$He} & {\bf $^7$Be(n,p)$^7$Li} & {\bf all THM} & \cite{Coc2012} \\ \hline \hline 

  $\frac{^6Li}{H}$$\times$10$^{14}$&   1.18269&   1.24548&   1.20147&   1.26633&   1.19128&   1.18139&     1.18269&  1.18268 & 1.27373 & 1.23 \\ \hline 
 
 $\frac{^9Be}{H}$$\times$10$^{19}$&  8.78250&   9.35568&   8.78744&   9.37543&   9.43551&   9.64746&     9.36744&  8.93604 & 11.92813 & 9.60 \\ \hline 
 
 
 $\frac{^{11}B}{H}$$\times$10$^{16}$&  3.25468&   2.90947&   3.22649&   2.87997&   2.85780&   3.56143&     3.25473&  3.03855 & 2.56751 & 3.05  \\ \hline 
 
 $\frac{CNO}{H}$$\times$10$^{16}$&  7.82870&   7.61318&   7.76462&   7.54749&    7.85210&   7.85350 & 7.93714& 7.82696 & 7.69346 & 7.43 \\ \hline
    \end{tabular}
  \caption{BBN abundances resulting from PRIMAT with only direct data reaction rates (first column), compared with the PRIMAT results using the THM rates one by one of the six reactions measured (see text). Ninth column is the results with the THM rates together, while the last one has the values of reference from calculations in \cite{Coc2012}.}
   \label{tab:calc}

\end{sidewaystable}

\section*{Conclusions}

The concept and applications of the THM, a powerful tool for investigating nuclear reactions at very low energies, were presented. This technique effectively allows to complement and strengthen direct measurement results in Nuclear Astrophysics.

For the first time, the overall impact of THM measurements on Primordial Nucleosynthesis has been evaluated with the PRIMAT code, to fix the model issues. 
The difference between light elements abundances calculated with the THM rates and with direct data lead to a better agreement with observational values. This highlights the relevance of obtaining more precise rates to confirm if THM rates can help in solving the $d-tension$ raised in \cite{anewtension} and the CLiP. 

Results for $^6$Li, $^9$Be and $^{11}$B isotopes are also closer to (or confirming for CNO isotopes) the expected values by \cite{Coc2012}, with respect to the case when only direct rates are used. 

In the light of these encouraging results, the extension of these calculation to the other THM measurement of BBN interest outside the main twelve reactions network (as the $^{7}$Be(n,$\alpha$)$^4$He \cite{Belicos}) and future THM measurements of other reactions of interest for SBBN (fig. \ref{network}), such as the $^3$H(d,n)$^4$He, can further improve the understanding of this crucial first period of nucleosynthesis, which sets the initial conditions for the following Universe evolution.

\section*{Acknowledgements}

This work was partially supported by the European Union (ChETEC-INFRA, project no. 101008324). The authors acknowledge "Piano Incentivi per la ricerca PIACERI 2024-26 " Linea di Intervento 1 “Progetti di ricerca collaborativa" by University of Catania.

\bibliography{biblio}

\end{document}